\documentclass[useAMS,usenatbib]{mn2e}
\usepackage{graphicx,amsmath,txfonts}

\newcommand{\mearth}{M_\oplus}
\title[Disruption of co-orbital resonances during orbital migration]{Disruption of co-orbital (1:1) planetary resonances during gas-driven orbital migration}
\author[A. Pierens,  S.N. Raymond]{A. Pierens $^{1,2},$  S.N. Raymond$^{1,2,3}$\\
$^1$Universit\'e de Bordeaux, Observatoire Aquitain des Sciences de l'Univers,
    BP89 33271 Floirac Cedex, France \\
$^2$CNRS, Laboratoire d'Astrophysique de Bordeaux,
     BP89 33271 Floirac Cedex, France\\
$^3$ NASA Astrobiology InstituteÕs Virtual Planetary Laboratory}
\date{Released 2012 Xxxxx XX}

\pagerange{\pageref{firstpage}--\pageref{lastpage}} \pubyear{2014}

\def\LaTeX{L\kern-.36em\raise.3ex\hbox{a}\kern-.15em
    T\kern-.1667em\lower.7ex\hbox{E}\kern-.125emX}
\begin{document}
\label{firstpage}
\maketitle
\begin{abstract}
Planets close to their stars are thought to form farther out and migrate inward due to angular momentum exchange with gaseous protoplanetary disks.  This process  can produce systems of planets in co-orbital (Trojan or 1:1) resonance, in which two planets share the same orbit, usually separated by 60 degrees.  Co-orbital systems are detectable among the planetary systems found by the Kepler mission either directly or by transit timing variations.  However, no co-orbital systems have been found within the thousands of Kepler planets and candidates.  Here we study the orbital evolution of co-orbital planets embedded in a protoplanetary disk using a grid-based hydrodynamics code.  We show that pairs of similar-mass planets in co-orbital resonance are disrupted during large-scale orbital migration.  Destabilization occurs when one or both planets is near the critical mass needed to open a gap in the gaseous disk.  A confined gap is opened that spans the 60 degree azimuthal separation between planets. This alters the torques imparted by the disk on each planet -- pushing the leading planet outward and the trailing planet inward -- and disrupts the resonance.  The mechanism applies to systems in which the two planets' masses differ by a factor of two or less.  In a simple flared disk model the critical mass for gap opening varies from a few Earth masses at the inner edge of the disk to 1 Saturn-mass at 5 AU.  A pair of co-orbital planets with masses in this range that migrates will enter a region where the planets are at the gap-opening limit.  At that point the resonance is disrupted.  We therefore predict an absence of planets on co-orbital configurations with masses in the super-Earth to Saturn mass range with similar masses.  
\end{abstract}
\begin{keywords}
accretion, accretion discs --
                planet-disc interactions--
                planets and satellites: formation --
                hydrodynamics --
                methods: numerical
\end{keywords}

\section{Introduction}
A striking feature of the current population of exoplanets is the broad diversity in system architectures that have been discovered.  
Among the known multi-planetary systems, a significant number contains bodies in resonant or near-resonant configurations.  For example, the two giant planets in the Kepler-9 system are trapped in 2:1 resonance (Holman et al. 2010). The formation of mean-motion 
resonances is generally explained by the convergent migration of two planets in the highly-damped environment of protoplanetary disks (e.g., Snellgrove et al. 2001; Lee \& Peale 2002; but see Raymond et al. 2008a).  

The origin of close-in systems of low-mass planets is debated (see Raymond et al 2008b, 2014).  Many of these systems' characteristics can be reproduced by either in-situ accretion within high-mass disks (Raymond et al. 2008b; Hansen \& Murray 2012, 2013; Chiang \& Laughlin 2013) or by accretion during inward migration of planetary embryos (Terquem \& Papaloizou 2007; Ogihara \& Ida 2009; Cossou et al. 2014).  The in-situ accretion model requires extremely massive inner disks with a broad range of surface density slopes, which is inconsistent with any known disk theory (Raymond \& Cossou 2014).  The inward migration model is viable but remains to be tested quantitatively.  There exist other plausible models that invoke large-scale inward migration of small bodies to produce a surplus in mass in the inner  parts of the disk (Boley \& Ford 2013; Chatterjee \& Tan 2014), essentially producing the initial conditions needed for the in-situ migration model.  Systems of low-mass planets are rarely in resonance.  Rather, pairs of adjacent planets are often found just wide of resonance (Lissauer et al. 2011).  It is unclear whether this supports the in-situ (Petrovich et al 2013) or migration (Baruteau \& Papaloizou 2013; Goldreich \& Schlichting 2014) model or whether it is due to another dynamical effect such as tides (Delisle et al. 2012; Lithwick \& Wu 2012; Batygin \& Morbidelli 2013).  

The inward migration model proposes that populations of embryos  migrate (Ward 1997; Tanaka et al. 2002) inward in so-called Type I migration from large orbital radii.  During this migration they interact gravitationally and occasionally merge (Terquem \& Papaloizou 2007; Ogihara \& Ida 2009; Cossou et al. 2014).  Using hydrodynamical simulations, Cresswell \& Nelson (2006, 2008) found that the formation of co-orbital planets engulfed in a 1:1 resonance is a natural outcome of close encounters between embryos during this process. Trojan planets are formed in $\sim 30 \%$ of their simulations.  Although they noted an increase  in the amplitude libration about the $L_4/L_5$ point during large-scale migration, Cresswell \& Nelson (2009)  confirmed that Trojan planet systems embedded in protoplanetary disks are stable. These studies concluded that co-orbital planets on short period orbits should be common.

Trojan planets could be discovered as two planets transiting the same star but separated by $60^\circ$ in phase.  They can also be detected by comparing the time of central transit with the time of zero stellar reflex velocity (Ford \& Gaudi 2006) or by transit timing variations (TTV; Ford \& Holman 2007).  Alternately, they can in principle be discovered within radial velocity datasets (Laughlin \& Chambers 2002).  These techniques -- in particular the TTV method -- are capable of detecting sub-Earth-mass co-orbital companions to gas giants.  However, to date no Trojan planets have been found in the {\it Kepler} data (Janson 2013). One reason for that may be related to the fact that only Trojans with coplanar orbits can be detected with {\it Kepler}. It is also possible that existing Trojans evolve on horseshoe or tadpole orbits with large libration, and thus are easily missed from planet-search algorithms since they induce large transit timing variations (Janson 2013).

In this paper we propose that the lack of detected Trojan planets is caused in part by the disruption of co-orbital systems during gas-driven migration.  
This destabilization occurs when both co-orbital planets can open a partial gap around their orbit (Lin \& Papaloizou 1993; Crida et al. 2006).  We use hydrodynamical simulations to show that the gap region located between the planets' orbits is far more gas-depleted than the rest of the disk. This essentially produces a gap that is confined in azimuth between the planets.  This makes the trailing planet experience a negative torque whereas the leading component feels a positive torque from the disk. The planets experience divergent migration and migrate divergently, resulting in the disruption of the 1:1 resonance. 
In an evolved, flared disk this mechanism destabilizes co-orbital systems with two Saturn-mass planets at several AU.  Trojan configurations with two $5-10 \mearth$ planets become unstable in the inner disk due to the ability of such planets to open a partial gap in the disk when the disk aspect ratio is small.  Co-orbital systems with more massive trailing planets are less stable than systems with more massive leading planets. This mechanism relies on non axisymmetric perturbations to the disk structure and requires that the two planets' masses be within roughly a factor of two.  Trojan systems with disparate masses or in the Jupiter-mass range are stable during migration.  
 
The paper is organized as follows. We describe the numerical set-up in Sect. 2. In Sect. 3, we present the physical mechanism responsible for the unstable evolution of co-orbital systems and examine the orbital migration of Trojan planets for a variety of planet masses and disk models.  Finally, we summarize and conclude in Sect. 4.

\section{Numerical set-up}
Simulations were performed using the GENESIS (De Val-Borro et al. 2006) numerical code which solves 
the equations governing the disk evolution on a polar grid $(R,\phi)$ using an advection scheme based on the monotonic  transport 
algorithm (Van Leer 1977). It includes a fifth-order Runge-Kutta integrator (Press et al. 1992) that computes 
the planet orbits.

The computational units we adopt are such that the mass of the central star $M_\star=1$ 
corresponds to one Solar mass, the gravitational constant is $G=1$ and the radius $R=1$ in the 
computational domain corresponds to $5$ AU. In the following, time is measured in units of the 
orbital period at $R=1$.

We use $N_R=768$ radial grid cells uniformly distributed between $R_{in}=0.25$ and 
$R_{out}=2.4$, and $N_\phi=1200$ azimuthal grid cells.  For a planet-to-star mass ratio $q=3\times 10^{-5}$, which 
corresponds to a $10$ Earth mass planet, and a disc aspect ratio $h=0.05$, the dimensionless half-width of the 
horseshoe region is $x_s\sim 1.2\sqrt{q/h}\sim 0.027$ (e.g. Paardekooper et al. 2010).  This means that the half-width 
of the horseshoe region is resolved by about $10$ cells in the radial direction. To avoid wave reflections at the disk edge, we employ 
wave-killing zones for $R>2.1$. At the inner edge, we use a viscous outflow boundary condition  
(Pierens \& Nelson 2008), where the radial velocity in the ghost zones is set to  $v_r=\beta v_r(R_{in})$, 
where $v_r(R_{in})=-3\nu/2R_{in}$ is the gas drift velocity due to viscous evolution and $\beta$ is a free parameter 
which was set to $\beta=5$.

 For most of the simulations presented here, we adopt a locally isothermal equation of state such that both the disk temperature and aspect ratio profiles are constant 
in time. Our fiducial disk model adopts a constant aspect ratio with  $h=0.05$ or $h=0.02$ but we also considered a flared disk model 
with $h=h_p(R/R_p)^\beta$, where $\beta=0.3$ is the flaring index and $h_p$ the disk aspect ratio at $R_p=5$ AU.  
We have also performed an additional set of simulations using a radiative disk model, in order to test 
the dependence of our results upon the choice of the equation of state. The disk viscosity is $\nu=10^{-5}$ in code units which corresponds to an alpha viscous stress parameter $\alpha=4\times 10^{-3}$ for $h_p=0.05$. The initial surface density profile is chosen to be $\Sigma=\Sigma_p(R/R_p)^{-3/2}$ with $\Sigma_p=2\times 10^{-4}$. 

The trailing and leading components of the co-orbital system are both initiated on circular orbits at $R_p=1$  and are initially 
located at their mutual $L_4/L_5$ Lagrange points. In the following, we denote by $m_t$ and $m_l$ the masses of the trailing and leading planets 
respectively, for which we consider values between $1.6 \mearth$ and $1 M_{J}$, where $M_{J}$ is the mass 
of Jupiter. 
\begin{figure}
\centering
\includegraphics[width=0.99\columnwidth]{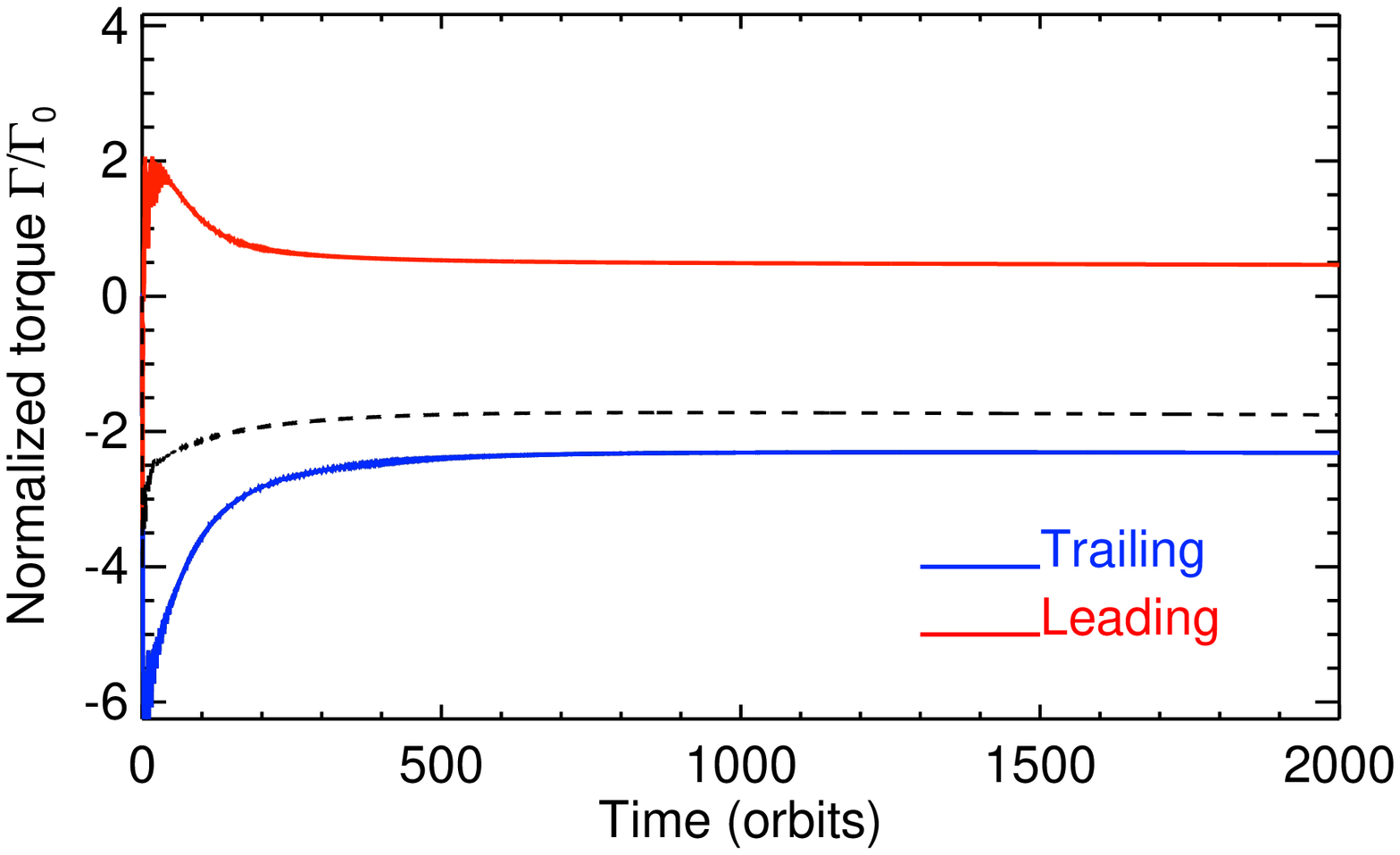}
\includegraphics[width=0.99\columnwidth]{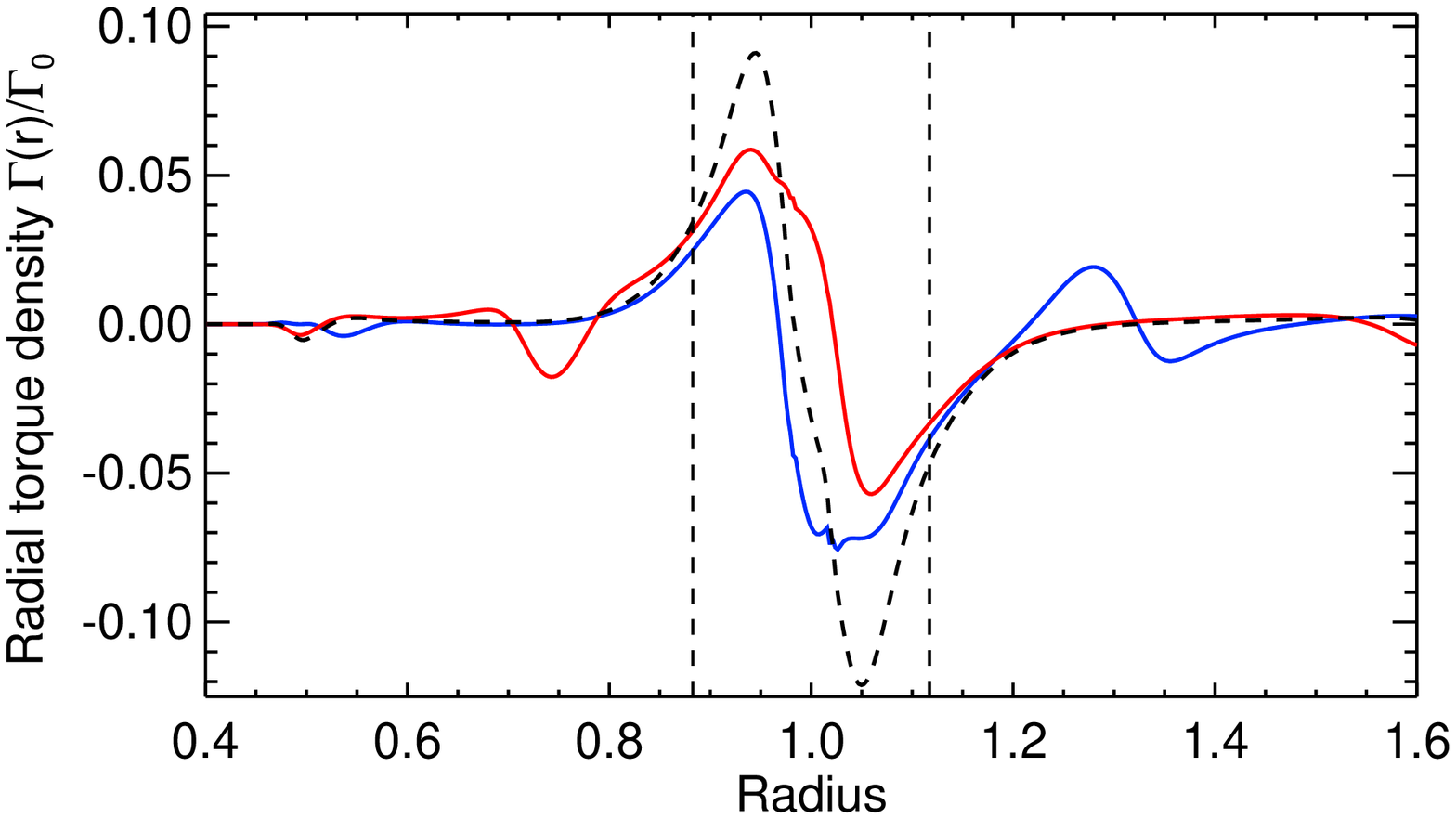}
\caption{{\it Upper panel}: Time evolution of the disk torques experienced by the trailing (blue) and leading (red) 
 components of a co-orbital system with $m_t,m_l=1\; M_S$.  The dashed line represents the torque evolution for a 
 single Saturn-mass planet.  {\it Lower panel:} Radial torque distribution for the 
 trailing and leading planets. The dashed line corresponds to the radial torque density distribution of a single  Saturn-mass planet   and the vertical dashed lines delimit the horseshoe region.}
\label{fig:torques}
\end{figure}
\section{Results}
\subsection{Disk torques exerted on the trailing and leading components}
\label{sec:torques}

We first examine the torques experienced by co-orbital planets held on fixed orbits.  We use a simple disk model with constant aspect ratio $h=0.05$. Fig. \ref{fig:torques} (upper panel) shows the evolution of the  torques  $\Gamma$ 
normalized by $\Gamma_0=(q/h)^2\Sigma_p R_p^4 \Omega_p^2$, where $\Omega_p$ is the planet angular 
velocity, experienced by the trailing and leading components of a co-orbital system with $m_{t},m_{l}=1$ $M_S$, where $M_S$ is the mass of Saturn. The disk torques exerted on the two co-orbitals are different, even though the planets share the same orbit.  The trailing planet feels a positive torque and the leading planet feels a negative torque.  This should lead to divergent orbital migration of the co-orbital pair.  To explain the differences in the magnitudes of the torques, we show the  torque density distribution acting on the trailing and leading components in the lower panel of Fig. \ref{fig:torques}. Significant differences arise from inside the horseshoe region of the planets, whose half-width for a Saturn-mass planet is estimated to be $x_s=2.45 R_H$ (Masset et al. 2006), where  $R_H=R_p(M_S/3M_\star)^{1/3}$ is the Hill radius of the planet. Compared with the torque exerted on a single planet (the dashed line in the lower panel of Fig. \ref{fig:torques}) the torque exerted on the leading planet takes more positive values over the whole horseshoe region whereas the torque acting on the trailing planet takes more negative values. Slight differences between the torques exerted on the two co-orbitals can also be observed outside of the horseshoe region. In the outer disk, the torque felt by the leading planet is in good agreement with the torque exerted on a single planet.  In the inner disk it is the torque felt by the trailing planet that matches the single-planet torque. This suggests that there is an additional positive contribution from the outer disk region bounded by $1.2<R<1.3$  to the torque exerted on the trailing planet; 
as well as an additional negative contribution from the inner disk region bounded by $0.8<R<0.9$ to the torque acting on the leading planet.  As shown below, the differences in the torque density distribution outside of the horseshoe region results from the interaction of the trailing (resp. leading) planet with the 
wake of the leading (resp. trailing) planet. However, the fact that the total torque exerted on the leading planet is positive suggests that these additional contributions to the torque density distribution are smaller than the differences in the torques arising from inside the horseshoe region.

\begin{figure*}
\centering
\includegraphics[width=0.325\textwidth]{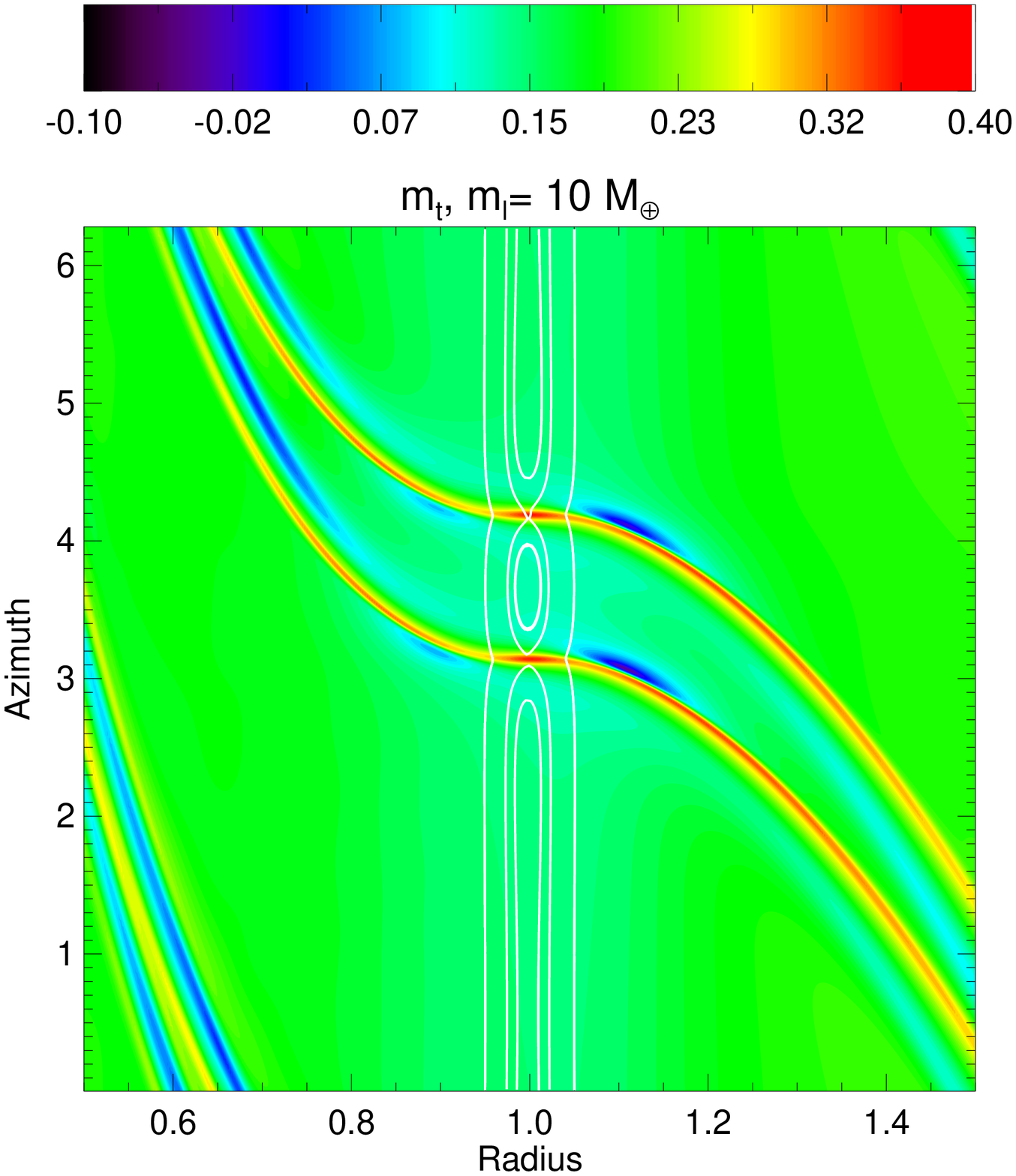}
\includegraphics[width=0.325\textwidth]{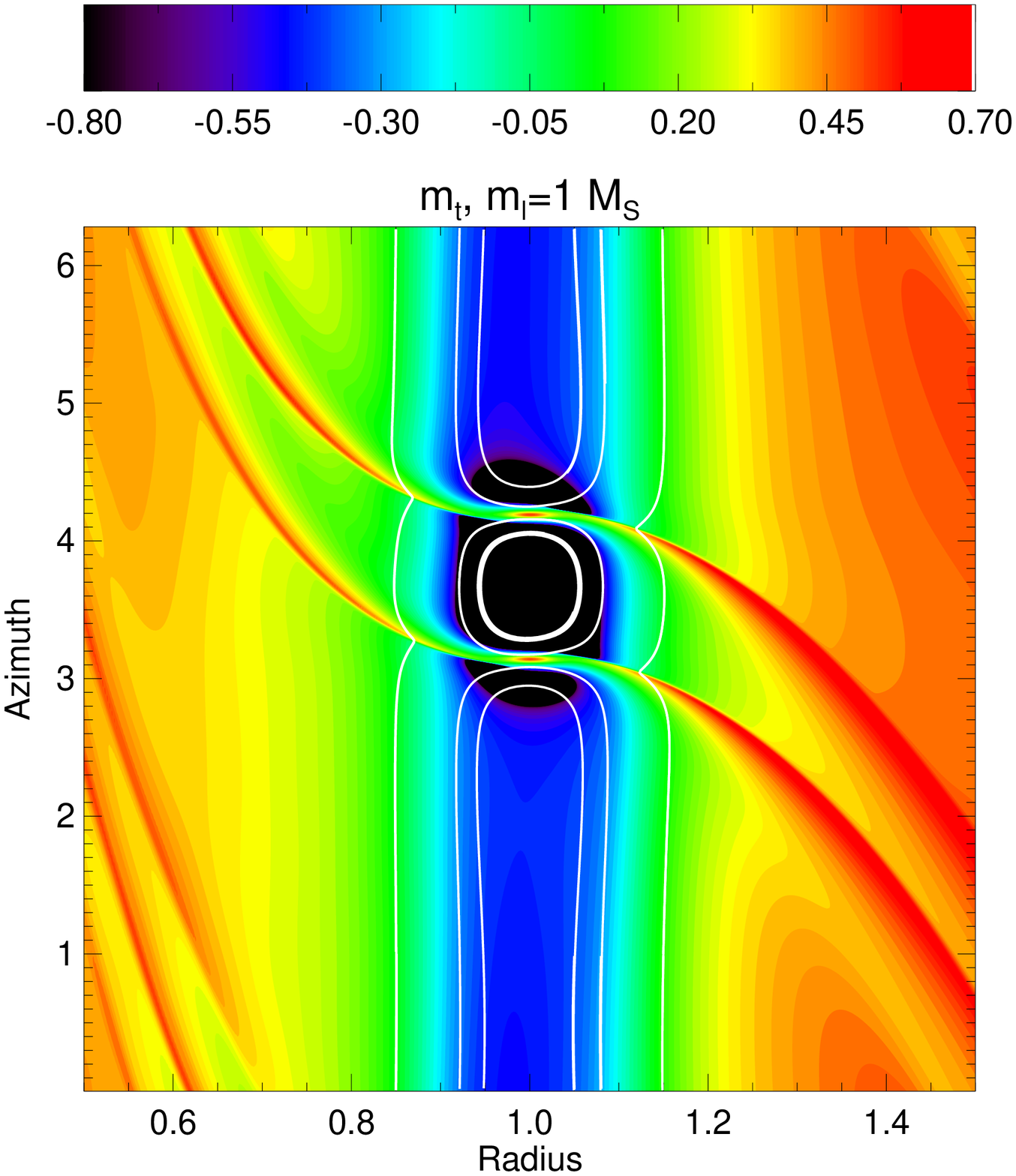}
\includegraphics[width=0.325\textwidth]{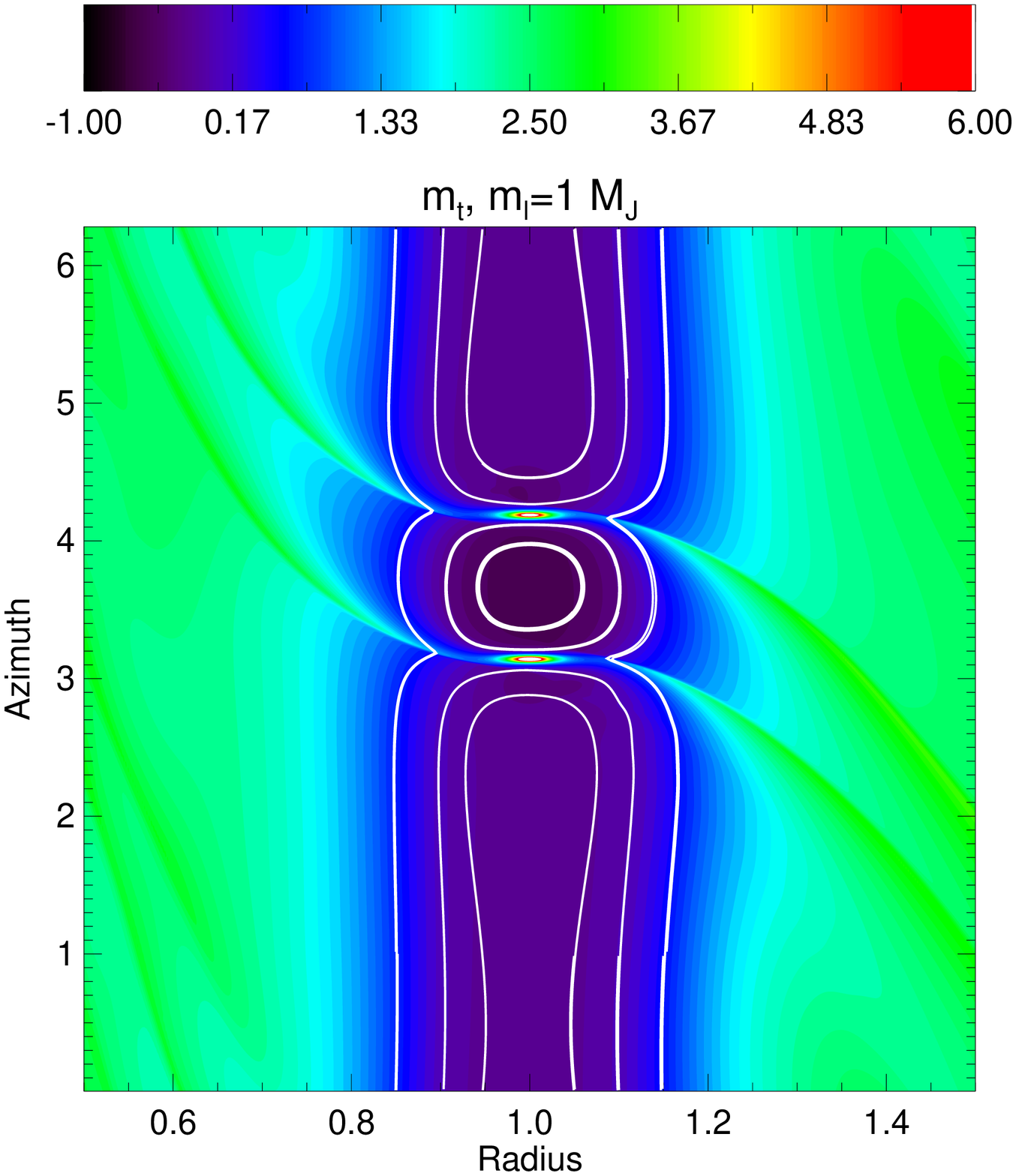}
\caption{Contours of the perturbed surface density $(\Sigma-\Sigma_0)/\Sigma_0$, 
where $\Sigma_0$ is the initial surface density profile,  for simulations with equal-mass co-orbitals of masses 
$m_t,m_l=10M_\oplus$ (left panel), $1M_S$ (middle panel), $1M_J$ (right panel). A few streamlines are 
overplotted as white lines.}
\label{fig:2dplots}
\end{figure*}

Fig. \ref{fig:2dplots} shows contours of the perturbed surface density   $(\Sigma-\Sigma_0)/\Sigma_0$, 
where $\Sigma_0$ is the initial surface density profile, for simulations with two equal-mass co-orbital planets with, from left to right, $m_t,m_l= 10 \mearth, 100 \mearth \approx M_S,$ and $300 \mearth \approx M_J$.  In the figure the trailing planet is located at $\phi_t=\pi$ rad and the leading one at $\phi_l=4\pi/3$ rad.  Overplotted are a few streamlines that delimit two different horseshoe regions, i) one of azimuthal extension $\Delta \phi=\pi/3$ rad and located 
ahead the trailing planet and behind the leading component and ii) one of larger azimuthal extension $\Delta \phi=5\pi/3$ and bounded by $\phi=\phi_l$ and $\phi=2\pi+\phi_t$. 

As expected, the density structure is not significantly perturbed for low-mass planets ($m_t,m_l=10\mearth$).  However, Trojan planets with $m_t,m_l>M_S$ produce gaps in the disk. The azimuthal density structure of the gap is strongly asymmetric in the simulation with $m_t,m_l=M_S$, with a much more gas-depleted region located ahead the trailing planet and behind the leading one.  The implication is that  in this case, the trailing (resp. leading) planet tends to feel a strong negative (resp. positive) corotation torque due the lower surface density ahead (resp. behind) of the planet.  This explains the torque density distribution in the lower panel of Fig. \ref{fig:torques}.
This figure also shows that the differences in the torque distribution outside of the horseshoe region are likely to 
result from the interaction between one component of the co-orbital system and the wake of the second component. In the outer disk, the wake of the leading planet and shearing past the trailing planet tends to exert a positive torque on the 
latter, clearly seen in the torque density distribution of the training planet from $R=1.2$ to $R=1.3$. Alternatively, 
the wake of the trailing planet and shearing past the leading component in the inner disk exerts a negative torque on it, 
in agreement with the torque distribution of the leading planet for $0.8<R<0.9$.
The asymmetry in the azimuthal gap structure in the middle panel of Fig. \ref{fig:2dplots} should occur for co-orbital planets that are able to open a partial gap in the disk. For a single planet on a circular orbit with semi-major axis $a_p$, the first condition for gap clearance is  $R_H>H$ (Ward 1997), where $H$ is the disk scale height, so that the wake of the planet forms a shock  and the flux of angular momentum carried by the wake is deposited locally. Gap formation also requires that the planetary tidal torque exceeds the viscous torque $J_{visc}$ which is given by (Lynden-Bell \& Pringle 1974):
\begin{equation}
J_{visc}=3\pi \nu \Sigma_p a_p^2 \Omega_p
\end{equation}
This condition leads to the so-called viscous criterion for gap-opening $q>40 \nu/a_p^2\Omega_p$ (Lin \& Papaloizou 1993). Crida et al. (2006) provided a single criterion for gap-opening which combines the two aforementioned conditions and  reads ${\cal P}<1$, where the gap-opening parameter $\cal{P}$ 
is given by:
\begin{equation}
{\cal P}=1.1\left(\frac{q}{h^3}\right)^{-1/3}+\frac{50\nu}{qa_p^2\Omega_p}
\label{eq:crida}
\end{equation}
We now turn to the issue of the conditions for gap clearance for two co-orbital planets located at their mutual 
$L_4/L_5$ Lagrange points.  The main difference with the single planet case is that the integrated viscous torque now depends on which part of the horseshoe region is considered. For the horseshoe region of 
 azimuthal extension $\Delta \phi=\pi/3$ and located ahead the trailing planet and behind the leading 
 body, the integrated 
viscous torque over the region is $J_{visc}(\Delta\phi=\pi/3)=J_{visc}/6$ whereas for the  part of the horseshoe region with azimuthal extension $\Delta \phi=5\pi/3$ and bounded by $\phi_l<\phi<2\pi+\phi_t$ , 
the integrated viscous torque is $J_{visc}(\Delta \phi=5\pi/3)=5J_{visc}/6$, where $J_{visc}(\Delta \phi=2\pi)=J_{visc}$. Consequently, it is straightforward to show that the gap-opening criterion of Crida et al. (2006)  can be written for the two 
regions:
\begin{eqnarray}
{\cal P}_1=1.1\left(\frac{q}{h^3}\right)^{-1/3}+\frac{1}{6}\frac{50\nu}{qa_p^2\Omega_p}<1 &  \text{for $\phi_t<\phi<\phi_l$}\\
\label{eqn1}
{\cal P}_2=1.1\left(\frac{q}{h^3}\right)^{-1/3}+\frac{5}{6}\frac{50\nu}{qa_p^2\Omega_p}<1  & \text{for $\phi_l<\phi<2\pi+\phi_t$}
\label{eqn2}
\end{eqnarray}
For $h=0.05$ and $\nu=10^{-5}$, Eq. 3 predicts that a gap is formed in this region for $m_t,m_l=M_S$ whereas from Eq. \ref{eqn2}, we find that  a mass of $m_t,m_l=M_J$ is required to clear a gap in the region bounded by $\phi_t$ and $2\pi+\phi_l$. In that case, and as illustrated by the right panel of Fig. \ref{fig:2dplots}, the azimuthal structure of the gap becomes axisymmetric such that the difference in the torques acting on the trailing and leading planets is weakened in comparison with the case with $m_t,m_l=M_S$. 

 The gap tends to be cleared more quickly in the region bounded by  $\phi_t<\phi<\phi_l$  than in the rest of the disk. Inspection of surface density maps at different times suggests that it takes approximately $\sim 10$ orbits to form the gap in the region in between the two planets while the gap is carved in $\sim 100$ orbits in the rest of the disk. Consequently, we expect that the asymmetry in the gap structure, and 
therefore the difference in the torques felt by the 
trailing and leading planets is maximal after $\sim 10$ orbits. The gap structure then becomes slightly more symmetric 
due to the formation of a shallow gap in the rest of the disk and the differences in torques weaken, consistent with the time evolution of the 
torques in Fig. \ref{fig:torques}.

\begin{figure} 
\centering
\includegraphics[width=0.99\columnwidth]{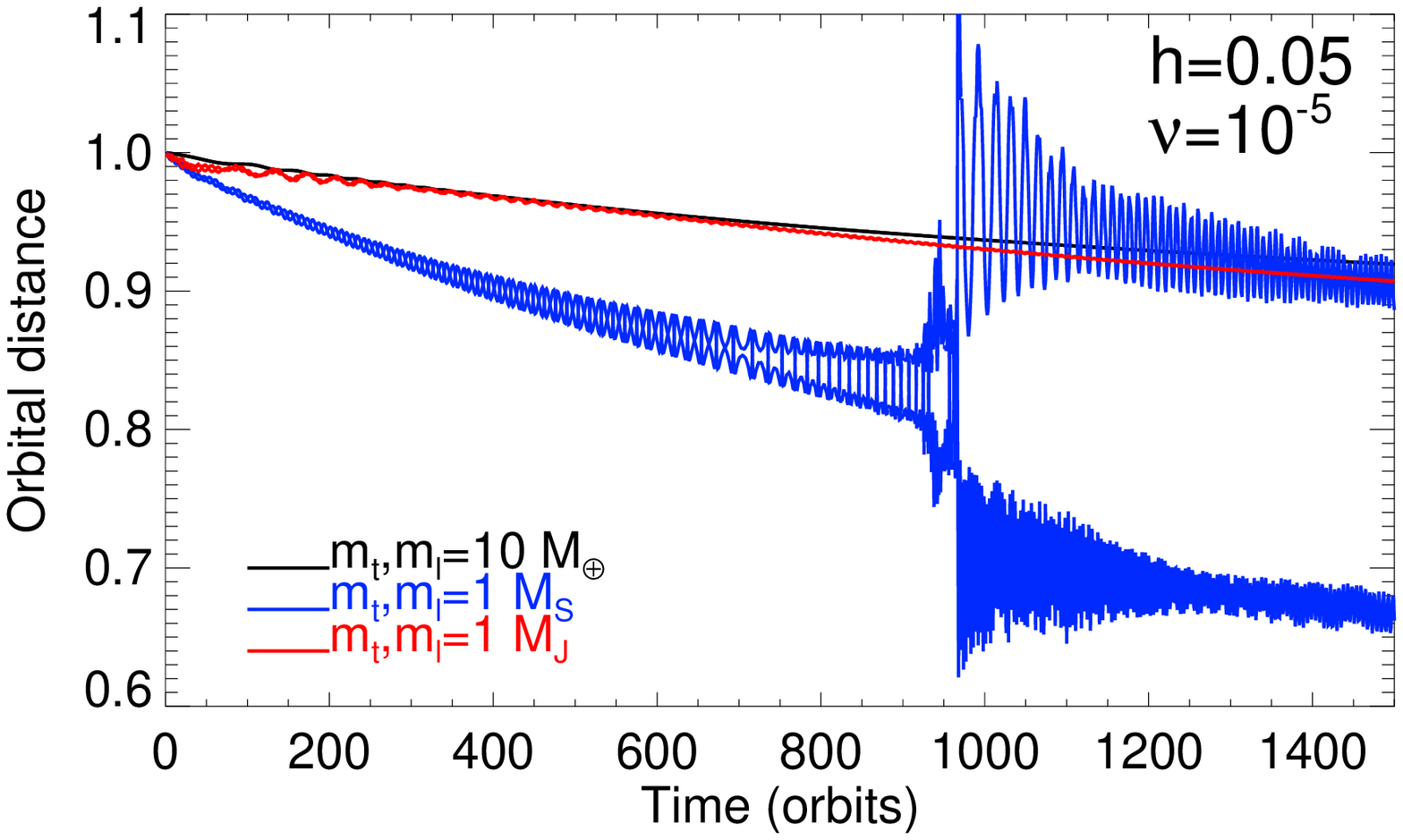}
\includegraphics[width=0.99\columnwidth]{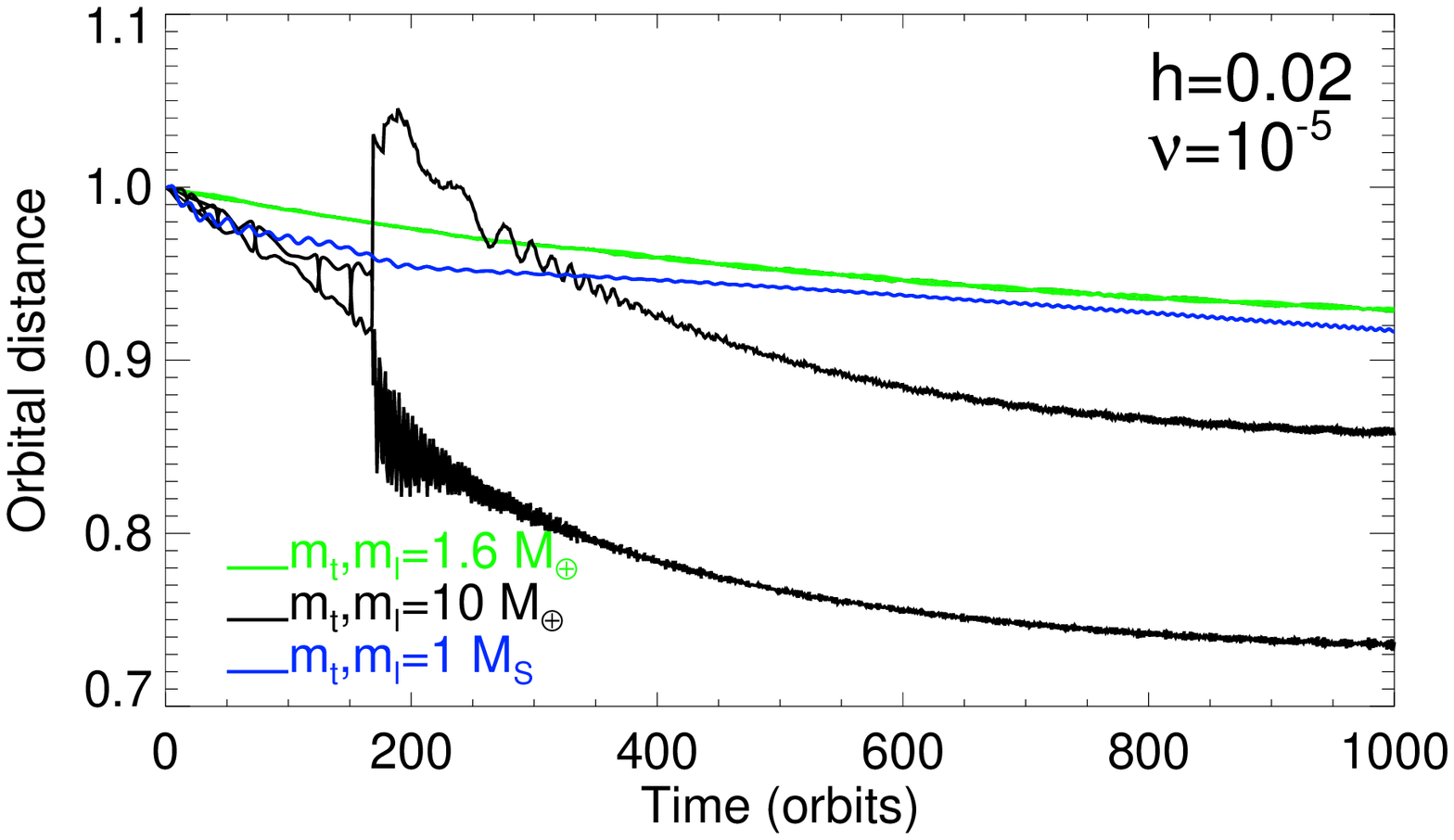}
\caption{{\it Upper panel:} Time evolution of the orbital distance of co-orbital systems with $m_t,m_m=10\mearth$ 
(black), $M_S$ (blue), $M_J$ (red) for a disk model with constant aspect ratio $h=0.05$. {\it Lower panel:} 
same but for $h=0.02$ and $m_t,m_l= 1.6\mearth$ (green), $10\mearth$ (black), and $M_S$ (blue).}
\label{fig:moveh5}
\end{figure}

\subsection{Stability of migrating co-orbital systems.}
\subsubsection{Disk with constant aspect ratio}

Fig. \ref{fig:moveh5} (top panel) shows the evolution of three co-orbital systems in a disk with aspect ratio $h=0.05$.  Each system has two equal-mass planets with masses of $10 \mearth$ (black curves), $100 \mearth (\approx M_S$; blue curves) and $300 \mearth (\approx M_J$; red curves).  A locally isothermal equation of state is adopted in these simulations such that all systems migrate inward.  For $m_t,m_l = 10 \mearth$
the orbital evolution of the trailing and leading components are almost indistinguishable, 
indicating that the libration amplitude about the $L_4/L_5$ Lagrange points is small and that the system 
is likely to be stable. Using a more realistic equation of state may lead to a different mode of evolution since it has been shown that in 
non-isothermal disks, the non-barotropic part of the corotation torque can make Type I migration slow down 
 or even reverse (Paardekooper \& Mellema 2006; Baruteau \& Masset 2008). The issue of the equation of 
 state is probably not important for the simulations with $m=M_S$ or $M_J$ in which the planets open a gap in the 
 disk and deplete their horseshoe region, weakening thereby the effect of the corotation torque 
  (Kley \& Crida 2008).  We will discuss 
 in more details the sensitivity of our results on the equation of state in Sect. \ref{sec:rad}.
 \begin{figure*}
\centering
\includegraphics[width=0.325\textwidth]{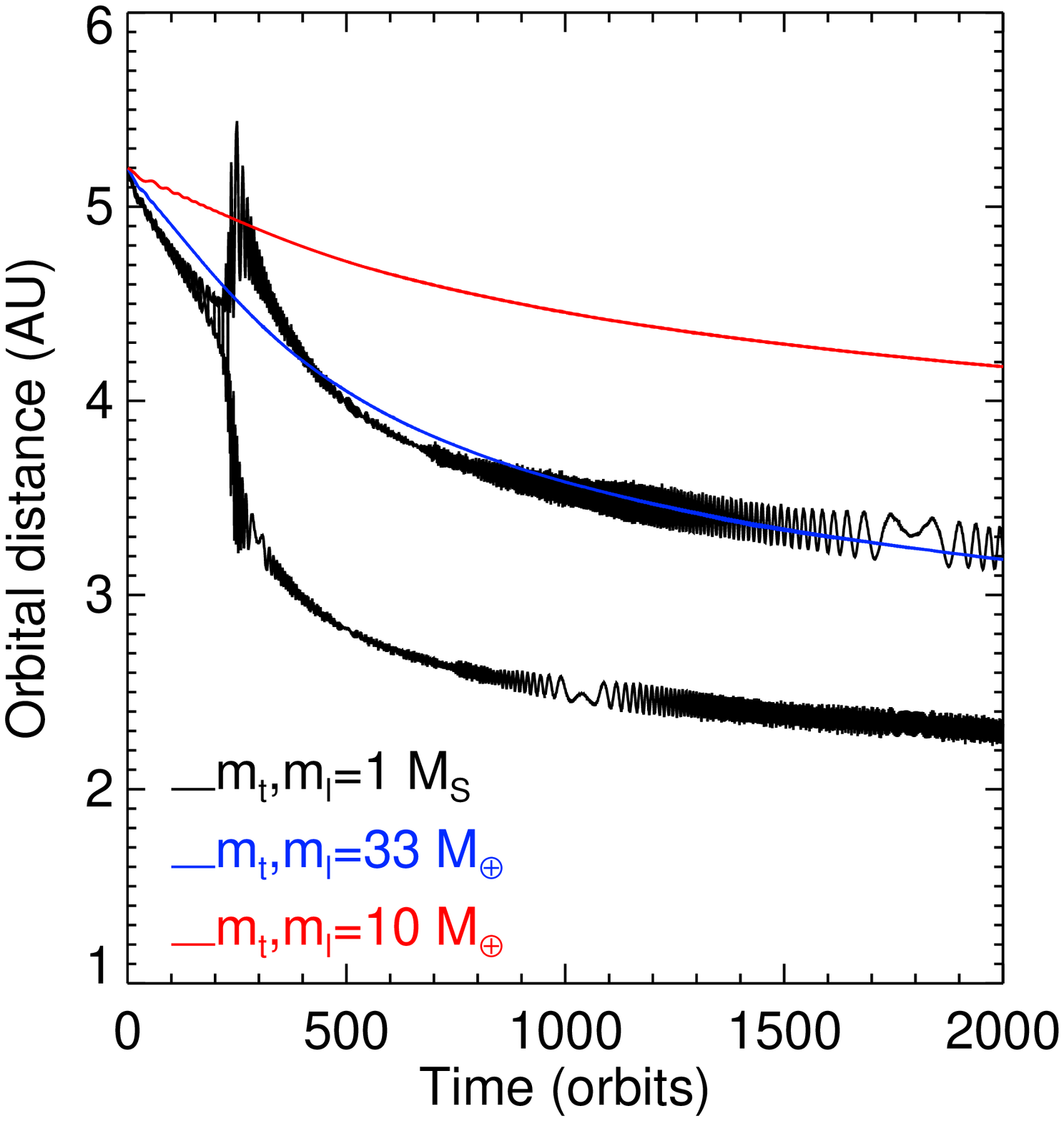}
\includegraphics[width=0.325\textwidth]{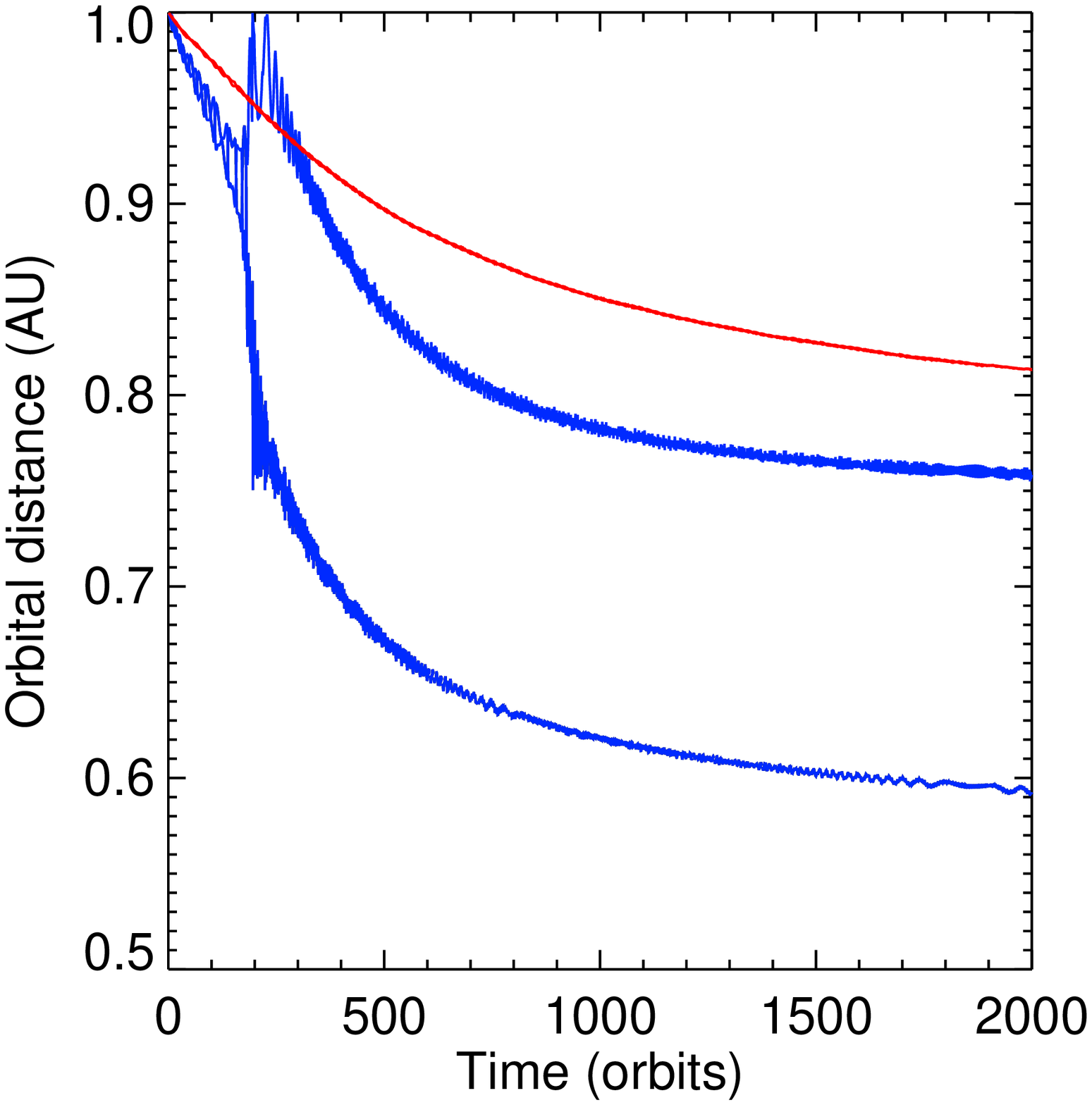}
\includegraphics[width=0.325\textwidth]{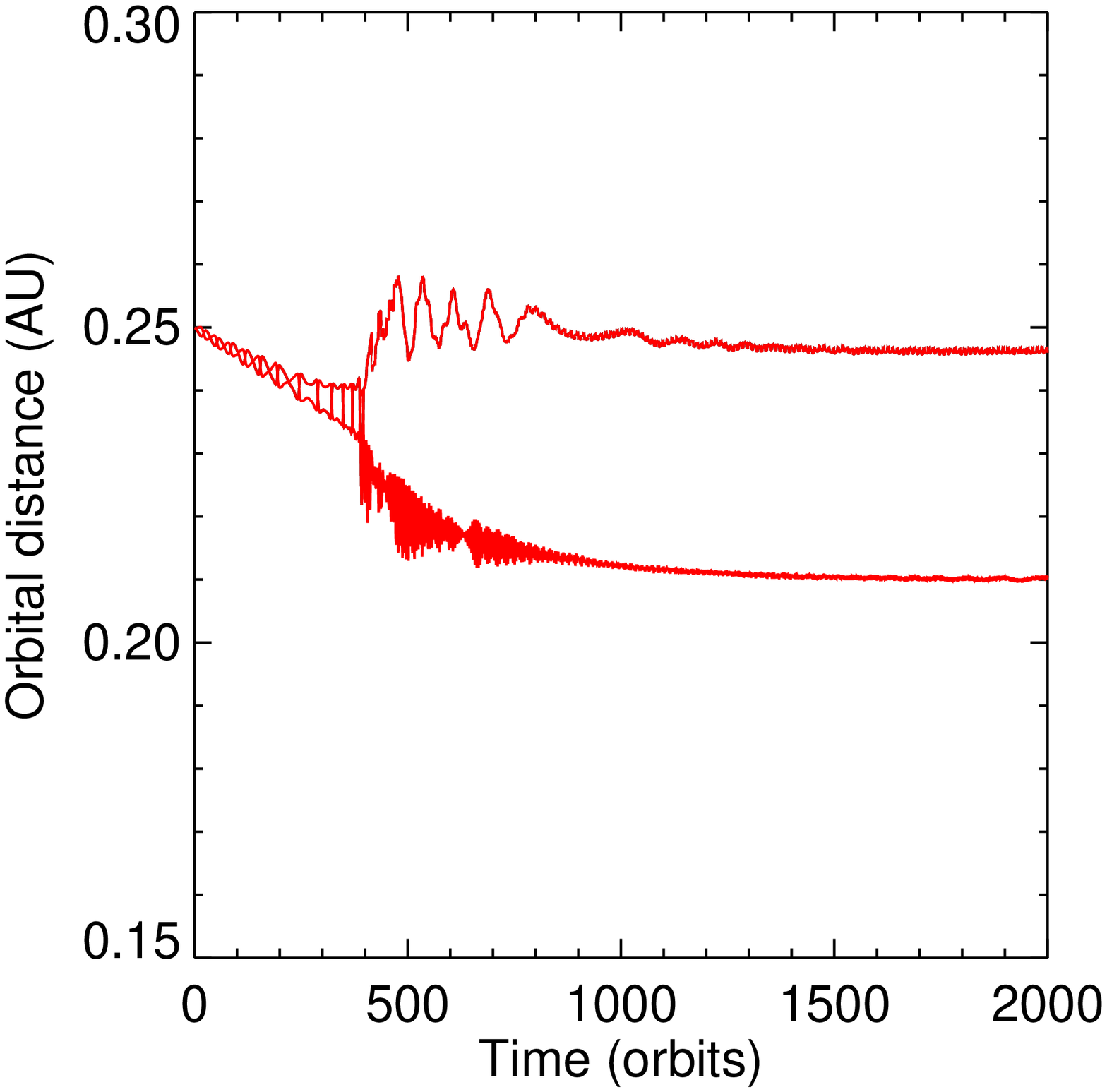}
\caption{Evolution of co-orbital systems with $m_t,m_l=M_S$ (black), $m_t,m_l=33 \mearth$ (blue), 
$m_t,m_l=10\mearth$ (red) in a flared disk model with $h=0.05(R/\text{5 AU})^{0.3}$. {\it Left panel:} time evolution 
of the planets' orbital distances in disk regions  $1<R<5$ AU. {\it Middle panel:} Same but for $0.5<R<1$ AU. {\it 
Right panel:} same but for $R<0.3$ AU.}
\label{fig:flared}
\end{figure*}
Different evolution occurs for more massive co-orbitals. Trojan planets with $m_t,m_l=M_J$ open a deep gap in the disk and undergo stable Type II migration.  However, the co-orbital configuration with $m_t,m_l=M_S$ is disrupted at $t\sim 1000$ orbits.  This occurs because the structure of the gap cleared by both planets is such that the leading planet feels a positive torque from the disk.  The trailing planet feels a negative torque, resulting in divergent migration. In the case with $m_t,m_l=M_J$ the surface density in the gap is only very slightly non-axisymmetric so the torques exerted on both planets are similar and the system is stable.

We now demonstrate that the mechanism presented in Sect. \ref{sec:torques} is responsible for breaking the co-orbital resonance for the case $m_t,m_l=M_S$.  We performed similar hydrodynamical simulations but for a disk model with $h=0.02$.  The co-orbitals that were destabilized should have masses in the partial-gap opening regime for which $q/h^3\sim 1$ (e.g. Baruteau \& Papaloizou 2013).  We therefore expect Trojan planets with masses typical of Super-Earths to become unstable when $h=0.02$.  Fig. \ref{fig:moveh5} (bottom panel) shows the orbital evolution for simulations with $m_t,m_l=1.6\mearth, 10\mearth$ and $M_S$.  As expected, the co-orbital system with $m_t,m_l=10\mearth$ is indeed unstable and the one with $m_t,m_l=M_S$ is stable. The perturbed density has an asymmetric gap structure to that in the middle panel of Fig. \ref{fig:2dplots}.  In this thin ($h=0.02$) disk the system with Saturn-mass co-orbitals carves a deep, axisymmetric gap and is stable.

\subsubsection{The case of a flared disk}

So far, we have considered a simplified disk with constant aspect ratio. In a more realistic model the aspect ratio is a sensitive function of the opacity and energy flux due to both viscous heating and stellar irradiation (Bitsch et al. 2013). For an evolved disk, we expect stellar heating to dominate over viscous heating and the disk to become flared with an aspect ratio $h\propto R^{2/7}$ (Chiang \& Goldreich 1997; Marzari \& D'Angelo 2012).  What is particularly interesting about a flared disk is that the partial gap-opening mass is a function of orbital radius.

To investigate the evolution of co-orbital systems embedded in a flared disk we performed an additional suite of simulations with $m_t,m_l=10\mearth, 33\mearth$ and $M_S$.  The disk model's aspect ratio is $h=h_p(R/R_p)^{0.3}$, where $h_p=0.05$ and $R_p$ is the initial position of the planets, set to be 5~AU. Due to the computational expense of the simulations, we study the large-scale migration of these systems using three different sets of simulations covering different radial zones: $1<R<5$ AU, $0.5<R<1$ AU and $0.15<R<0.5$ AU.

Fig. \ref{fig:flared} (left panel) shows the orbital evolution of co-orbital systems for $1<R<5$ AU.  As for a constant aspect ratio with $h=0.05$, the co-orbital configuration with $m_t,m_l=M_S$ is unstable once the planets reach $R_p\sim 4.5$ AU.  At this location, the planets are in the partial gap-opening regime, with $q/h_p^3\sim 2$.  The planets open a non-axisymmetric gap since ${\cal P}_1\sim 1$ and ${\cal P}_2\sim 2$. For $m_t,m_l=10\mearth$ ($m_t,m_l=33 \mearth$),  $q/h_p^3\sim 0.2$ ($q/h_p^3\sim 0.8$) so that the planets do not open a gap and their evolution is stable.  

As the planets migrate inward, the disk's local aspect ratio continuously decreases and we expect these systems to carve a gap in the disk and to become unstable once $q/h_p^3\sim 1$.  This is indeed what happens.  The co-orbital system with $m_t,m_l=33 \mearth$ is unstable at $\sim 0.9 AU$ (middle panel of Fig. \ref{fig:flared}) where the aspect 
ratio is $h_p \sim 0.03$ ($q/h^3\sim 3.5$). At this point,  ${\cal P}_1 \sim 1.5$ and ${\cal P}_2 \sim 5$ which 
confirms that the gap is strongly non axisymmetric in that case.  The system with $m_t,m_l=10\mearth$ is unstable at $\sim 0.23 AU$ (right panel of Fig. \ref{fig:flared}) where $h\sim 0.02$ ($q/h^3\sim 3.7$). At this location, we note that  
${\cal P}_1 \sim 3$ and ${\cal P}_2 \sim 15$.

Our results suggest that in a flared disk, equal-mass co-orbital systems become unstable where the planets can clear a partial gap. This corresponds to Saturn-mass planets located in the giant planet formation region outside the snow-line $R\sim 2.7$ AU (Lecar et al. 2006) or co-orbital planets of a few Earth masses in the inner regions of the disk $R\lesssim 1$ AU.  A co-orbital system can remain stable during migration if the two components are $\sim$Jupiter-mass and carve deep gaps in the disk.  Lower-mass co-orbitals could remain stable in a dead-zone where the viscosity is small enough for low-mass planets to open deep gaps and undergo Type II migration (e.g. Matsumura \& Pudritz 2007).

\section{Discussion}
Here we  discuss how our results depend on certain physical parameters.  We focus on 
the effect of the equation of state of the disk, the disk's viscosity, and the mass ratio between the trailing and leading planets.  We also examine the possible fates of planet pairs after disruption of the co-orbital resonance.
\subsection{Effect of the equation of state}
\label{sec:rad}

The simulations presented so far used a locally isothermal equation of state.  The disk's aspect ratio 
was constant. Although this is reasonable when modeling the outer, optically thin, regions of 
protoplanetary disks (see, e.g., Fig. 20 in Pierens \& Raymond 2011), this approximation breaks down in the optically thick inner parts where the equation of state is more likely to be adiabatic. To test the effect of the equation of state on our results we performed an additional set of simulations  using a non-isothermal disk model. The energy equation that we use includes the contribution from viscous heating plus a 
radiative cooling term $Q^-_{rad}=2\sigma_B T_{eff}^4$, where $\sigma_B$ is the Stefan-Bolzmann constant and 
$T_{eff}$ the effective temperature which is computed using the opacity law of Bell \& Lin (1994). For these calculations, 
the viscosity is $\nu=10^{-5}$ and the disk surface density at $R=1$ was chosen such that the disk aspect ratio 
is $h_p\sim 0.05$ at this location. 

\begin{figure} 
\centering
\includegraphics[width=0.99\columnwidth]{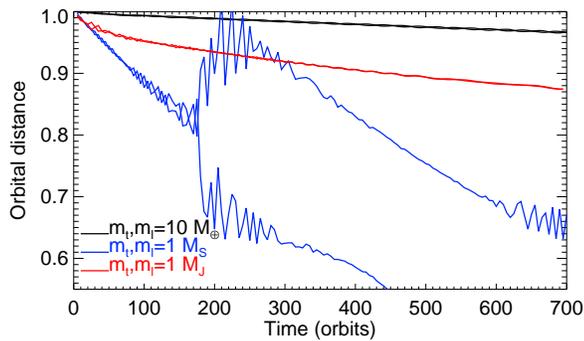}
\caption{Time evolution of the orbital distance of co-orbital systems with $m_t,m_m=10\mearth$ 
(black), $M_S$ (blue), $M_J$ (red) for the radiative disk model.}
\label{fig:rad}
\end{figure}

The results of these simulations are presented in Fig. \ref{fig:rad}.  The figure shows the evolution of the orbital distance for $m_t,m_l=10 M_\oplus, 1 M_S, 1M_J$. We note that for the disk parameters employed here, a simulation performed with a single $10$ $M_\oplus$ planet resulted in outward migration, which is consistent with the presence of  a negative entropy gradient in the disk (Baruteau \& Masset 2008).  The fact that a pair of $10\;M_\oplus$ co-orbital 
planets is observed to migrate inward is possibly related to the saturation of the corotation torque in the region 
located in between the planets, where the libration period is very short. In that case, only half of the corotation torque remains, which is clearly not enough to 
counterbalance the effect of the differential Lindblad torque. Comparing 
Fig. \ref{fig:rad} and the upper panel of Fig \ref{fig:moveh5} -- which corresponds to a locally isothermal equation 
of state with $h=0.05$ -- we see that very similar evolution outcomes are obtained. In both cases, we indeed find that the co-orbital resonance 
for $m_t,m_l=1 M_S$ is broken whereas the 1:1 resonance for $m_t,m_l=10M_\oplus$ or $m_t,m_l=1 M_J$ remains stable.  The 1:1 resonance is disrupted more quickly for the radiative disk model simply because the disk mass 
is higher in that case. This suggests that our results are fairly robust against the choice of the equation of state.

\subsection{Effect of the viscosity}

\begin{figure} 
\centering
\includegraphics[width=0.99\columnwidth]{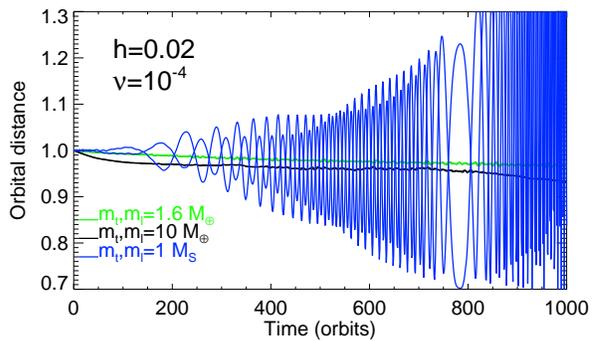}
\caption{Time evolution of the orbital distance of co-orbital systems with $m_t,m_m=1.6\mearth$ 
(green), $10\mearth$ (black), $M_S$ (blue) for an isothermal disk model with $h=0.02$ and $\nu=10^{-4}$.
 The case with $\nu=10^{-5}$ is shown in the bottom panel of Fig. \ref{fig:moveh5}.}
\label{fig:visc}
\end{figure}

As discussed earlier, the mechanism responsible for the destabilization of the 1:1 resonance occurs for planets 
in the partial gap opening regime.  This typically requires $q/h^3\sim 1$.  According to Eqs. \ref{eqn1} and $5$, 
we expect our results to depend on the value of the viscosity. For an inviscid disk,
the gap structure tends to be axisymmetric and therefore gap opening co-orbitals are stable, whereas for high 
values of the disk viscosity, the planets do not open a gap which prevents the mechanism to operate. In order to examine the 
effect of varying the disk viscosity, we performed a series of simulations for an isothermal disk model 
with $h=0.02$ and $\nu=10^{-4}$.  We remind the reader that in the case where $\nu=10^{-5}$, we found that 
$\sim 10\mearth$ co-orbitals become unstable due to the non axisymmetric depletion of the horseshoe region. 
Considering the case with
$\nu=10^{-4}$, we have $({\cal P}_1\,{\cal P}_2)\sim (30, 140)$ for $m_t,m_l=10\mearth$ whereas $({\cal P}_1,{\cal P}_2)\sim (3,14)$ for $m_t,m_l=1 M_S$, so that the  gap tends to be strongly non axisymmetric in the latter case while 
$10\mearth$ co-orbitals are not expected to open a gap  for $\nu=10^{-4}$. For this disk model, the orbital distance versus 
time for co-orbital planets with $m_t,m_l=1.6 \mearth$, $10\mearth$, $1M_S$ is plotted in Fig. \ref{fig:visc}. In agreement with the previous expectation, we indeed find that $10\mearth$ co-orbitals are now stable whereas the 1:1 resonance 
between planets with $m_t,m_l=1 M_S$ is much more chaotic, with the planets' eccentricity reaching values of 
$e_p\sim 0.4$.  Over longer timescales, the co-orbital pair is found to migrate outward due to the 
high value reached by the eccentricity, which is still growing .  Although the evolution outcome for this run remains uncertain over the timescale covered by the simulation, it seems likely that this system will ultimately become unstable.

\subsection{Evolution of co-orbitals of different mass}
We now discuss how the stability of the 1:1 resonance depends on the mass ratio $q_c=m_t/m_l$ between the trailing and leading planets. To investigate this issue, we have performed a suite of runs in a disk with constant aspect ratio $h=0.05$ in which the trailing or the leading planet is a Saturn-mass planet while the mass of the second component is varied in the range $[10\mearth, M_J]$. 

Fig. \ref{fig:varyq} shows the outcomes of the simulations.  In the simulations in the upper panel $q_c<1$ (the leading planet is more massive) and in the lower panel all simulations have $q_c>1$ (the trailing planet is more massive).  There is a clear tendency for the 
1:1 resonance to be more stable with a more massive leading planet ($q_c<1$).  For $q_c<1$ only one system is unstable, with $m_t=M_S$ and $m_l=1.7 M_S$.  But for $q_c>1$, the simulation with corresponding masses ($m_t=1.7 M_S$ and $m_l=M_S$)  was unstable on a much shorter timescale and two additional simulations were unstable, with $m_t=M_S$ , $m_l=66 M_\oplus$ ($q_c=1.5$) and $m_t=0.7 M_J$, $m_l=1 M_S$ ($q_c=2.3$). Co-orbitals systems with higher values of $q_c$ tend to be stable due to the ability of the more massive component to create a deep gap in the disk.

Co-orbital systems with more massive leading planets are more stable because when the two planets open a partial gap in the disk, the positive (resp. negative) contribution to the torque exerted on the leading (resp. trailing) component due to the more gas-depleted disk region located behind  (resp. ahead of)  this planet counterbalances the effect of a stronger (resp. smaller) negative differential Lindblad torque. In the case where the trailing planet is more massive, the contribution from the gap region to the torque exerted on the trailing planet and the differential Lindblad torque add, resulting in amplified differences in torques felt by the leading and trailing planets.

\subsection{Possible fates of  planet pairs after disruption of the 1:1 resonance}

\begin{figure}
\centering
\includegraphics[width=0.99\columnwidth]{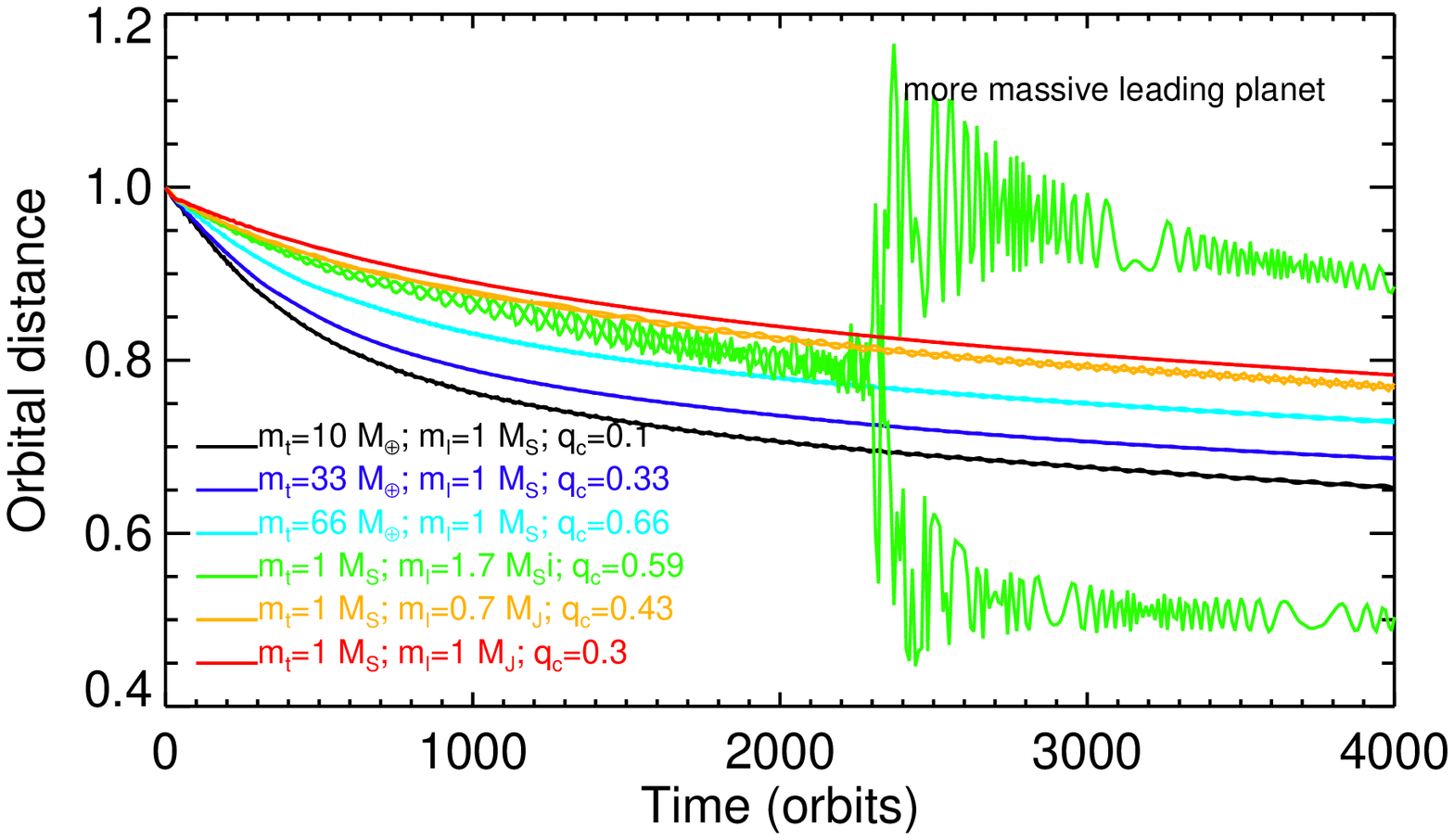}
\includegraphics[width=0.99\columnwidth]{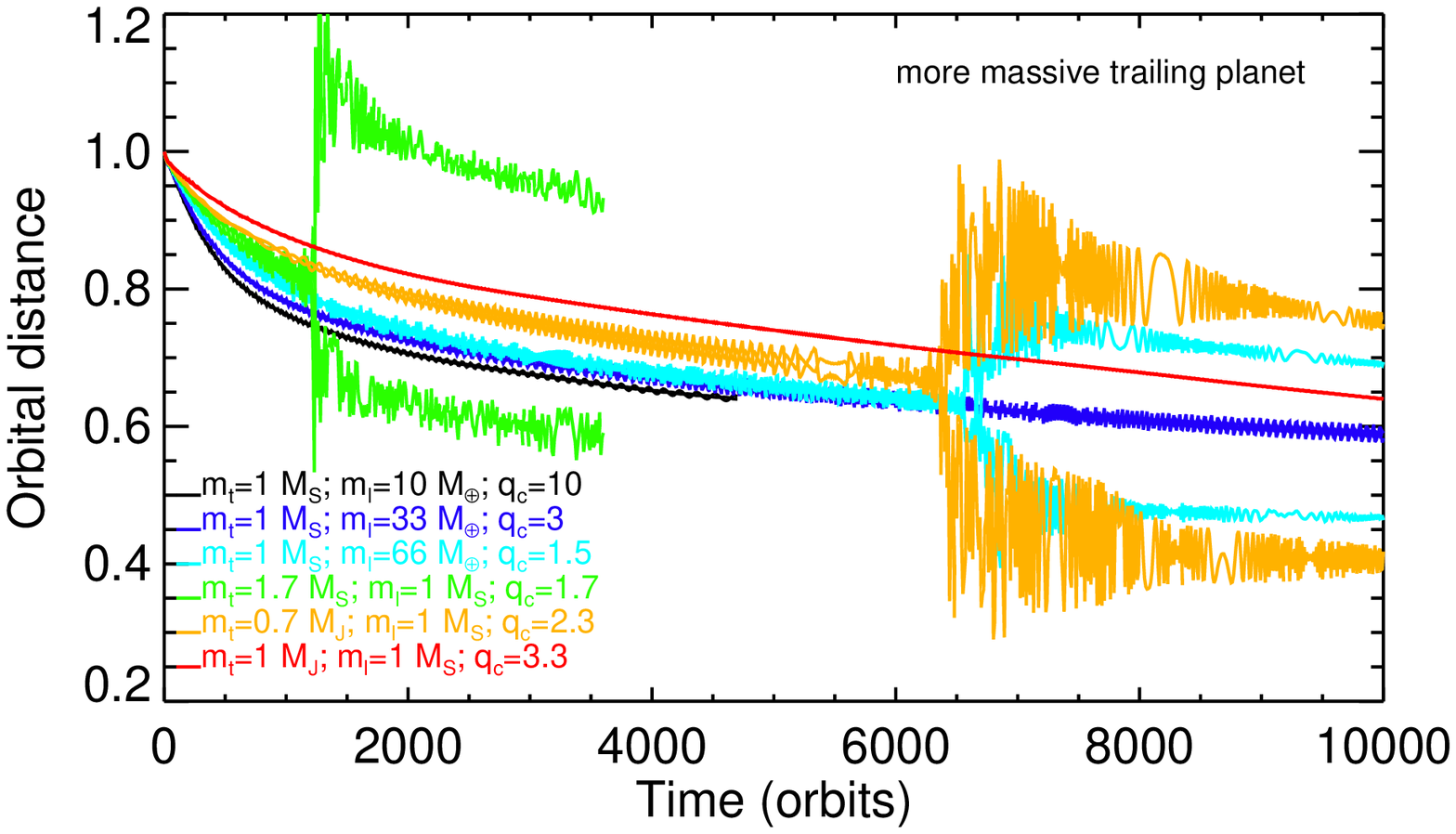}
\caption{{\it Upper panel:} time evolution of the orbital distances of Trojan planets for systems in which the leading 
component is the more massive ($q_c=m_t/m_l<1$). {\it Lower panel:} same but for co-orbital systems with a more 
massive trailing component ($q_c>1$). }
\label{fig:varyq}
\end{figure}

  For the isothermal simulations in which the 1:1 resonance is unstable, the long-term evolution outcome remains uncertain but it principle, systems in which the most massive planet is ejected to a larger radius may be trapped in a wider,  
 mean motion resonance,  since the planets tend to undergo convergent migration 
after disruption of the co-orbital resonance in that case.  For example, we checked that  the period ratio is continuously decreasing 
 after disruption of the co-orbital resonance in the runs of Fig. \ref{fig:varyq} with $q_c=1.5$ and $q_c=2.3$, 
 which suggests that  the planets may eventually become  trapped in a MMR. 
 An alternative possibility is that the planets become locked in a co-orbital resonance again.  It has been indeed shown that co-orbital planets can be formed in 
 isothermal disks during 
 the relaxation of a swarm of low-mass planets migrating inward  (Cresswell \& Nelson 2006).
 
  In the case of  radiative disk models, capture in a new 1:1 resonance may also occur for unstable systems with 
  equal-mass planets, for example during the convergent migration of protoplanets toward a zero-torque radius. 
  Formation of co-orbital planets was indeed observed to arise in the radiative simulations of Pierens et al. (2013), in which 
  $3M_\oplus$ embryos migrate toward a convergence line created by a change in the opacity regime.

\section{Conclusion}

We have used hydrodynamical simulations to study the orbital evolution of co-orbital planets located at their mutual $L_4/L_5$ Lagrange points embedded in a protoplanetary disk.  Co-orbital (also called Trojan or 1:1 resonant) configurations are disrupted when the planets open a partial gap around their orbit. This occurs because the gap that opens between the two planets is far more depleted than the rest of the co-orbital region (Fig. 2).  The trailing planet feels a negative torque and the leading planet feels a positive torque, resulting in divergent migration of the two planets' orbits. For a constant disk aspect ratio of $h=0.05$ and a viscosity typical to that in the active regions of protoplanetary discs, this mechanism destabilizes co-orbital systems in the Saturn-mass range (Fig. 3, top panel).  For a thinner disk with $h=0.02$ the 1:1 resonance is destabilized for planets of a few Earth masses (Fig. 3, bottom panel). Evolved protoplanetary disks are expected to be flared due to heating from the central star such that the partial gap-opening mass is a function of the orbital radius.  As they migrate inward, co-orbital configurations with different masses are therefore disrupted at different orbital radii.  Saturn-mass co-orbitals are disrupted beyond the snow line, $30\mearth$ co-orbitals are disrupted at $\sim 1$~AU and $10 \mearth$ co-orbitals are disrupted at a few tenths of an AU (Fig. 4). 
 
 Although most of our simulations were performed with an isothermal equation of state, a series 
of runs using radiative disk models indicated that our results are very robust regarding the choice of the 
equation of state. This occurs because the mechanism presented here is at work for planets that partially 
deplete their horseshoe region, and for which the effect of the corotation torque is substantially weakened.

In the partial gap-opening regime, co-orbital configurations are more stable in systems with more massive leading planets.  This is because the positive contribution to the torque from the gap region exerted on the leading component is balanced by a stronger (negative) differential Lindblad torque. For systems with a more massive trailing planet, the co-orbital system becomes unstable if $q_c=m_t/m_l\lesssim 2.5$. Systems with higher mass ratios are more stable due to the ability of the more massive component to create a deep gap in the disk. 

As they migrate inward, co-orbital systems can remain stable under certain conditions.  First, if both planets are in the full gap-opening regime (e.g., Jupiter-mass planets at 5~AU) then the gap is axisymmetric and too deep for this mechanism to operate.  Second, if the planets have a mass ratio larger than roughly a factor of two, then the perturbed surface density in the co-orbital region is nearly axisymmetric and the resonance remains stable.  

We therefore predict that no close-in co-orbital systems will be discovered with near equal-mass planets with masses between a few Earth-masses and a Saturn-mass.  

\section*{Acknowledgments}
Computer time for this study was provided by the computing facilities MCIA (M\'esocentre de Calcul Intensif Aquitain) of the Universite de Bordeaux and by HPC resources of Cines under the allocation c2013046957 made by GENCI (Grand Equipement National de Calcul Intensif).  We thank the Agence Nationale pour la Recherche under grant ANR-13-BS05-0003 (MOJO).  S.N.R.'s contribution was performed as part of the NASA Astrobiology Institute's Virtual Planetary Laboratory Lead Team, supported by the NASA under Cooperative Agreement No.  NNA13AA93A.

\end{document}